\documentclass{aa}  

\usepackage{graphicx}
\usepackage{txfonts}

\begin{document} 

\title{Temperature and density profiles in the corona of main-sequence stars induced by stochastic heating in the chromosphere}
   \titlerunning{Temperature and density profiles in the corona of main-sequence stars}
\author{Luca Barbieri
          \inst{1,2,3}
          \and
          Lapo Casetti
          \inst{1,2,3}
          \and
          Andrea Verdini
          \inst{1,2}
          \and
          Simone Landi
          \inst{1,2}  
          }
   \institute{
         Dipartimento di Fisica e Astronomia, Universit\`{a} di Firenze, via G.\ Sansone 1, 50019 Sesto Fiorentino, FI, Italy\\
         \email{luca.barbieri@unifi.it}
     \and
     INAF-Osservatorio Astrofisico di Arcetri, Largo E.\ Fermi 5, 50125 Firenze, Italy
     \and      
     INFN, Sezione di Firenze, via G.\ Sansone 1, 50019 Sesto Fiorentino, FI, Italy
             }
   \date{Received XX; accepted XX}
  \abstract 
{
All but the most massive main-sequence stars are expected to have a rarefied and hot (million-Kelvin) corona like the Sun. How such a hot corona is formed and supported has not been completely understood yet, even in the case of the Sun. Recently, \cite{Barbieri2023temperature,Barbieri2024b} introduced a new model of a confined plasma atmosphere and applied it to the solar case, showing that rapid, intense, intermittent and short-lived heating events in the high chromosphere can drive the coronal plasma into a stationary state with temperature and density profiles similar to those observed in the solar atmosphere. In this paper we apply the model to main-sequence stars, showing that it predicts the presence of a solar-like hot and rarefied corona for all such stars, regardless of their mass. However, the model is not applicable as such to the most massive main-sequence stars, because the latter lack the convective layer generating the magnetic field loop structures supporting a stationary corona, whose existence is assumed by the model. We also discuss the role of stellar mass in determining the shape of the temperature and density profiles.
}
   \keywords{Stars: coronae -- Stars: atmospheres -- Sun: corona -- Sun: atmosphere -- plasmas -- methods: analytical}

   \authorrunning{L.\ Barbieri et al.}
   
   \maketitle
\section{Introduction}
\label{sec:intro}
The density of the Sun steadily decreases from its center to the outermost layer of its atmosphere, the corona, which is extremely rarefied, being more than one million times less dense than the surface (photosphere). On the contrary, the temperature of the Sun decreases only up to the first layers of the atmosphere above the photosphere, and then starts to increase reaching millions of Kelvins in the corona, which is therefore more than two hundred times hotter than the photosphere \citep{GolubPasachoff:book}. This temperature increase while density decreases, often referred to as ``temperature inversion'', happens quite abruptly. Most of both the temperature jump and the density drop occur in the so-called ``transition region'', a thin (only hundreds of kilometers wide) layer separating the lowest layer of the Sun's atmosphere, the chromosphere, from the corona. Most of the plasma in the corona is organized in loops, following the magnetic field lines which exit the photosphere and then re-enter it, and emits radiation in the X band \citep{Aschwandensolarcorona}. An X-ray emission has been detected in many other main-sequence stars, regardless of the spectral type \citep{Pallavicini1989,Gdel2004,Xraycoronae}. Since all the main-sequence stars with mass $M < 1.5\, M_\odot$ have a convective region below the photosphere where a Sun-like magnetic field originates \citep[see e.g.][]{Maoz:1142636}, the X-ray emission is thought to be of coronal origin and all the late-type (from late spectral type A onwards) main-sequence stars are thought to have a corona analogous to the solar one, organized in loop structures and with a temperature around $10^6$ K. A strong hint at the presence of a corona made up of Sun-like magnetic structures is that the relation between the X-ray luminosity and the magnetic field strength follows the same power law in both the Sun's active regions \citep{Fisher_1998} and the atmospheres of less massive main-sequence stars \citep{Pevtsov_2003}. 
The case of the most massive main-sequence stars (i.e., O, B and up to early-A spectral type) is different because these stars lack a convective region below the photosphere and therefore should not have a solar-like magnetic field able to support a stationary corona: the X-ray emission from these stars is not attributed to a coronal activity but rather to strong winds and shocks \citep{Pallavicini1989,Gdel2004}.

Despite decades of investigation, the mechanism producing the million-Kelvin corona of the Sun (and of the other stars as well) is largely not yet understood. This open problem is referred to as the coronal heating problem \citep{Klimchuk_2006,2012coronalheating}. Most efforts have been conducted along the lines of finding suitable mechanisms able to transport energy from the lower layers to coronal heights, or to release energy stored in the magnetic field, and to efficiently dissipate such an energy in the corona \citep{Parker:1972wu,Dmitruk:1997uf,2005ApJ...618.1020G,Rappazzo:2008vl,2013ApJ...773L...2R,2015RSPTA.37340265W,Heyvaerts_Priest_1983,Ionson_1978,2020A&A...636A..40H,Pontieu:2011vg}. \citet{Scudder1992a,Scudder1992b} proposed a different approach, observing that collisions in the very dilute coronal plasma are rare and therefore the corona might be out of thermal equilibrium. Thus, if the velocity distribution functions of the particles at the base of the corona, i.e., in the upper chromosphere, are non-Maxwellian, and in particular have suprathermal tails (i.e., the probability of finding fast particles is larger than in a Maxwellian), then faster particles are able to climb higher in the gravitational potential well. As a result, the temperature increases with height while the density decreases. Such a mechanism was dubbed ``velocity filtration'' or ``gravitational filtering''.  At variance with the above-mentioned approaches, Scudder's model does not involve any local deposition of heat in the corona and is able to reproduce coronal temperatures and densities; however, it predicts a smooth change of the latter quantities, without a transition region, and its basic assumption of non-Maxwellian distribution functions for particles in the collisional chromosphere is difficult to justify.

In the highly collisional environment of the Sun's chromosphere, any deviation from thermal distributions is expected to be extremely short-lived. Therefore, particles' distribution functions in the chromosphere are expected to be thermal. This notwithstanding, the chromosphere is a very dynamic environment and its temperature is expected to fluctuate in space and time \citep{Molnar_2019}. Starting from this observation, \citet{Barbieri2023temperature,Barbieri2024b} recently reconsidered Scudder's pioneering intuition, replacing the (hardly justifiable) assumption of non-thermal distributions in the high chromosphere with the hypothesis of a fluctuating temperature in an otherwise fully collisional and thermal chromosphere. \citet{Barbieri2023temperature,Barbieri2024b} showed that rapid, intense, intermittent and short-lived temperature increments in the high chromosphere are able to drive the above plasma atmosphere towards a stationary configuration with an inverted temperature-density profile, with a transition region and a hot corona very similar to the observed one \citep{observedtemperature}, but for the fact that the transition region predicted by the model is thicker than the observed one. According to this model, the mechanism producing temperature inversion in the solar atmosphere is velocity filtration as in Scudder's model, but there is no need to postulate distribution functions with suprathermal tails in the chromosphere: the latter are self-consistently produced by the gravitational filtering itself, and originate in the superposition of different thermal distributions in the chromosphere. Temperature fluctuations are modelled by a stochastic process, and essentially any probability distribution such that more intense fluctuations are less frequent than less intense ones works. Remarkably, no fine-tuning of the parameters is required to produce temperature inversion, the only requirement being that the temperature fluctuations are fast enough to prevent the system from relaxing towards a thermal configuration (for the Sun, this means that fluctuations must occur on a subsecond time scale, which is unresolved in current solar observations). The average intensity of the fluctuations, however, must be sufficiently large as to produce coronal temperatures. Indeed, short-lived, intense, and small-scale brightenings are routinely observed on the Sun \citep{Dere:1989ux,Teriaca:2004wy,Peter:2014uz,Tiwari:2019us,Berghmans:2021wl}. The so-called campfires recently observed in extreme UV images have temperatures of the order of  $10^6$ K, while explosive events appearing in $\mathrm{H}\alpha$ line widths have smaller temperatures, about $2 \times 10^5$ K, but are ten times more frequent \citep{Teriaca:2004wy}. This trend is consistent with the distribution of rapid temperature increments assumed in the model. Moreover, based on recent extreme UV solar observations, \citet{Rauoafi:ApJ2023} have shown that small-scale magnetic reconnection events at the base of the solar corona can produce a flow of matter that propagates up into the corona with the correct energy budget to heat the plasma environments up to million degrees. 

Since a Sun-like hot corona is expected to be present in low-mass main-sequence stars  (i.e., stars with $M < 1.5\, M_{\odot}$), and the velocity filtration mechanism is very general and does not require any Sun-specific feature, in the present paper we apply the model proposed by \citet{Barbieri2023temperature,Barbieri2024b} to main-sequence stars. More specifically, we want to answer the following questions: Does the model predict an inverted temperature-density profile with a transition region and a hot corona for all main-sequence stars? How are these profiles affected by the mass of the star?

The paper is structured as follows. In Sec.\ \ref{sec:model} we briefly review the plasma atmosphere model proposed by \citet{Barbieri2023temperature,Barbieri2024b}: more specifically, in Sec.\ \ref{subsec:model_description} we describe the model and define its parameters, and in Sec.\ \ref{subsec:transition_region} we discuss the mechanism according to which the model produces a transition region and a hot corona, defining a quantity that discriminates the range of parameters in which a transition region and a corona do actually form from those in which they do not, and which will be useful in the following. In Sec.\ \ref{sec:main_sequence_stars_LMS}  we apply the theoretical framework to main-sequence stars and we present and discuss the results, focusing on the case of low-mass stars ($M < 1.5\, M_{\odot}$; the case of higher-mass stars is considered in Appendix \ref{sec:main_sequence_stars_HMS}). 
Finally, in Sec.\ \ref{sec:conclusions} we summarise our findings and hint at some possible follow-ups of the present work. 

\section{The model}
\label{sec:model}
\subsection{Model description}
\label{subsec:model_description}
\begin{figure}
    \centering
    \includegraphics[width=0.99\columnwidth]{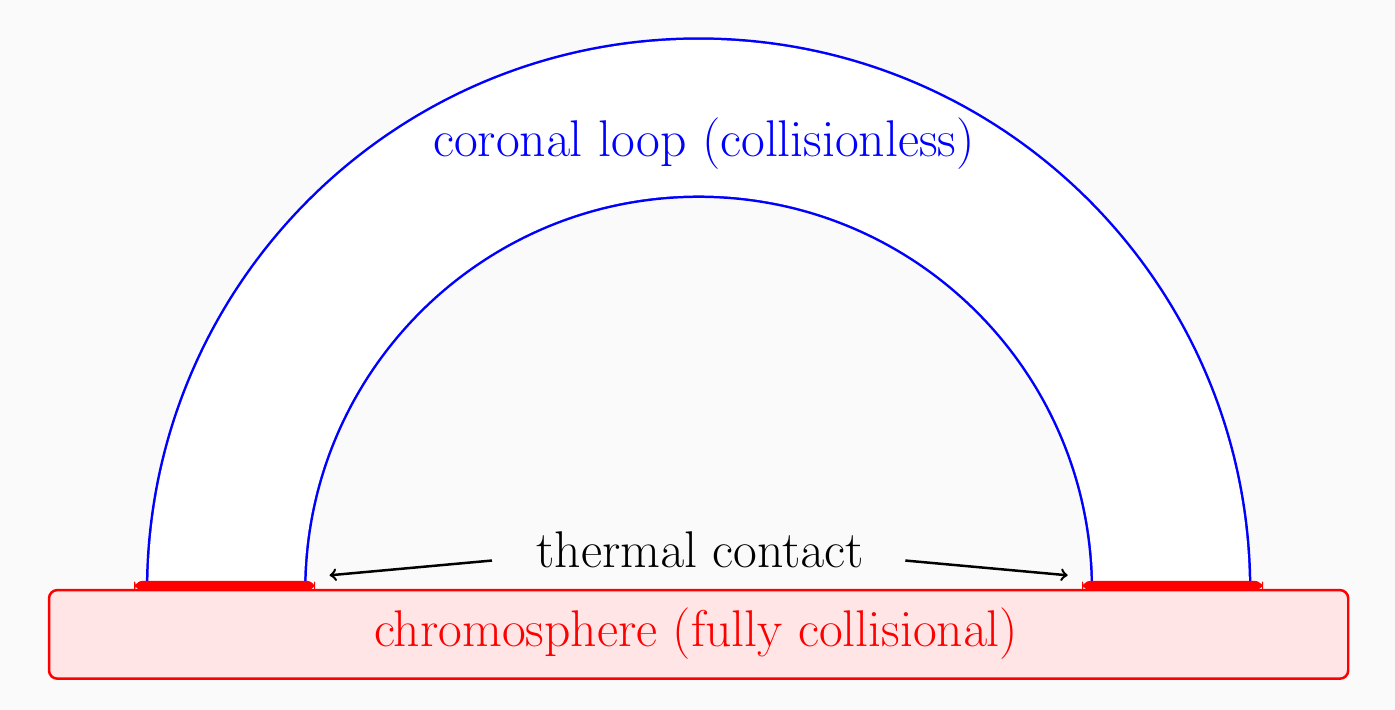}
    \caption{Schematics of the loop model. The coronal plasma in the loop is treated as collisionless and in thermal contact with a fully collisional chromosphere (modeled as a thermostat).
    }
    \label{fig:Loopscheme}
\end{figure}

\begin{figure}[b!]
    \centering
    \includegraphics[width=0.99\columnwidth]{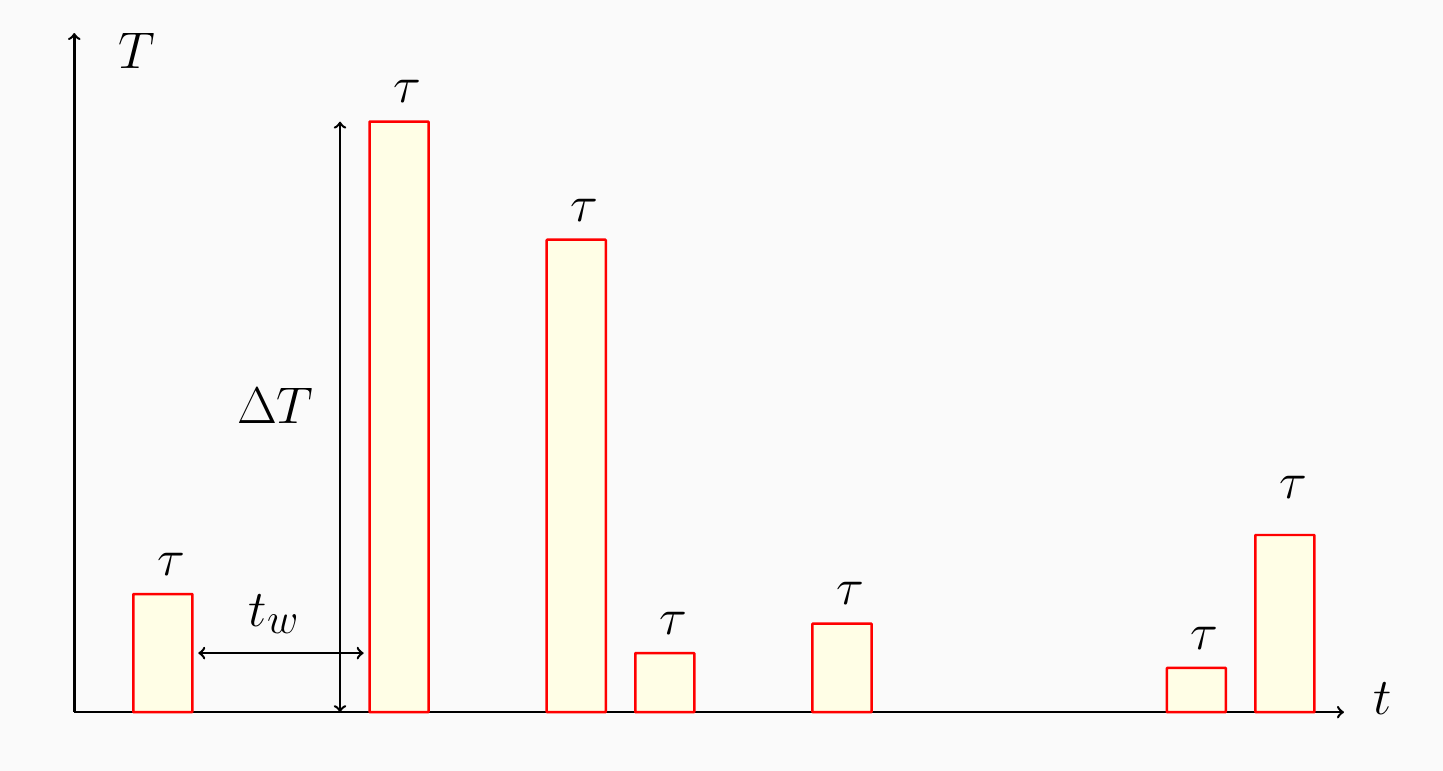}
    \caption{Sketch of the time series of the temperature of the thermostat (chromosphere). During the time intervals of duration $\tau$ the temperature increases by an amount $\Delta T$ and during the waiting times $t_w$ it returns to the typical chromospheric value $T_b$.
    }
    \label{fig:Temperatureflucscheme}
\end{figure}

Here we briefly summarise the model of the plasma atmosphere introduced by \citet{Barbieri2023temperature,Barbieri2024b}. 
A coronal loop in the atmosphere of a star is modeled as a semicircular tube of length $2L$ made up of a two-component (electrons and protons), collisionless electrostatic plasma subjected to an external force field consisting of a constant gravity plus an electric field ensuring charge neutrality and proportional to $g(m_e+m_p)/2$, where $m_e$ and $m_p$ are the masses of the electrons and of the ions, respectively \citep{Pannekoek_1922,Rosseland_1924,belmont2013collisionless}, and $g=GM/R^2$ is the surface gravity with $M$ and $R$ the mass and radius of the star, respectively. Particles are allowed to move only along the loop. This structure is assumed to be in ideal thermal contact with a thermostat that mimics the fully collisional chromosphere. A scheme of the model is shown in figure \ref{fig:Loopscheme}.
If the temperature $T_b$ of the thermostat (i.e., of the chromosphere) is constant, the coronal loop is in thermal equilibrium at the same temperature of the thermostat. However, as mentioned in the Introduction, the chromosphere is a very dynamic environment; therefore, we assume that its temperature fluctuates due to heating events of amplitude $\Delta T$ and duration $\tau$, separated by waiting times $t_w$ during which the temperature of the thermostat switches back to $T_b$. A sketch of the time series of the temperature of the thermostat is depicted in figure \ref{fig:Temperatureflucscheme}.
We keep $\tau$ fixed and we draw $\Delta T$ and $t_w$ from probability distributions. If the following conditions are fulfilled,  
\begin{equation} \label{eq:time_constraints}
    \tau < t_r \quad \text{and} \quad \langle t_w \rangle_\eta < t_r,
\end{equation}
where $t_r$ is the relaxation time in the corona and $\langle \cdot \rangle_{\eta}$ is the average over a given probability distribution $\eta$, then the coronal plasma never relaxes to thermal equilibrium: it rather reaches a stationary state always exhibiting inverted temperature and density profiles. 
The relaxation time $t_r$ can be estimated as the minimum between the thermal crossing time $t_T$ and the free-fall crossing time $t_g$ of the electrons, i.e.,
\begin{equation}\label{relaxationtimes}
        t_r = \text{min}(t_T;t_{g})\,,
\end{equation}
where
\begin{equation}\label{relaxationtimes2}
        t_T=2L \sqrt{\frac{m_e}{k_B \langle T \rangle_{\gamma}}} \quad \text{and} \quad
        t_{g}=\sqrt{\frac{2 L m_e}{g(m_e+m_p)}}~.
\end{equation}
In Eq.\ \eqref{relaxationtimes2} above, $\langle T \rangle_{\gamma}$ is the average of $T$ over a given probability distribution $\gamma$ of temperature increments and $k_B$ is the Boltzmann constant.

Since the coronal plasma dynamics is treated as collisionless, the time evolution of the distribution functions of both species obeys Vlasov equations \citep[see e.g.][]{nicholson1983introduction}.
By time averaging the Vlasov dynamics and the fluctuating thermostat over a timespan long enough to encompass many temperature increments, it is possible to analytically calculate the full phase-space distribution functions in the steady state (for details see \citealt{Barbieri2024b}). The resulting expressions for the single-particle phase space distribution functions $f_e$ and $f_p$ of electrons and ions, respectively, are
\begin{equation} \label{eq:falphastationary}
    f_{\alpha}(x,v) = \mathcal{N}_{\alpha} \left[ (1-A)\frac{e^{-\frac{H_{\alpha} (x,v)}{k_B T_{b}}}}{T_{b}}+A\int_{T_b}^{+\infty} dT\, \gamma(T)\, \frac{e^{-\frac{H_{\alpha} (x,v)}{k_B T}}}{T}\right]\,,
\end{equation}
where $\alpha = e$ or $p$, the constant $A$ is given by 
\begin{equation}\label{eq:A}
     A = \frac{\tau}{\tau + \langle t_w \rangle_{\eta}},
\end{equation}
and $H_{\alpha}$ is the single particle energy of the species $\alpha$, i.e.,
\begin{equation}\label{eq:singleparticlehamiltonian}
    H_{\alpha}=\frac{m_{\alpha} v^2}{2}+g\frac{m_e+m_p}{2}\,z 
\end{equation}
where $v$ is the particle velocity and 
\begin{equation}\label{eq:def_z}
    z=2\frac{L}{\pi}\cos{\left(\frac{\pi x}{2L}\right)}
\end{equation}
is the height upon the surface at a given position corresponding to a curvilinear abscissa $x \in [-L,L]$ along the loop. 

The constants $\mathcal{N}_{\alpha}$ are normalisation constants, so that the distribution functions $f_{\alpha}$ are normalised to $1$. 

The interpretation of Eq.\ \eqref{eq:falphastationary} is as follows: the distribution functions in the stationary state are given by a thermal distribution at temperature $T_b$ (the reference temperature of the thermostat) plus a non-thermal contribution resulting from the average of thermal distributions at a temperature $T>T_b$ over the probability distribution $\gamma(T)$ of the temperature increments. The weight of the non-thermal contribution is proportional to $A$, which is the fraction of time the thermostat is not at temperature $T=T_b$, i.e., the fraction of time in which the chromosphere is actually heated. The thermal population dominates at small heights $z$ and is suppressed by the gravitational term in $H_\alpha$ as $z$ increases; conversely, the non-thermal contribution becomes increasingly relevant at larger $z$ due to velocity filtration, as faster particles can climb higher in the potential well and appear as suprathermal tails in the distribution functions. 

To perform the calculations, we choose the distributions of the temperature increments, $\gamma$, to be an exponential distribution,  
\begin{equation}\label{eq:prob_increments}
     \gamma(T)=\frac{1}{\langle \Delta T\rangle}e^{-\frac{T-T_b}{\langle \Delta T \rangle}}, \quad T>T_b\,, 
 \end{equation}
where $\langle \Delta T \rangle = \langle T - T_b \rangle_\gamma$ is the mean value of the temperature increments.

This is the simplest choice guaranteeing that large temperature increments are less likely than small ones, as suggested, for instance, by the fact that the so-called ``campfires'' recently observed in extreme UV solar imaging have temperatures of about $10^6$ K, while explosive events appearing in $\mathrm{H}\alpha$ line widths have smaller temperatures, about $2 \times 10^5$ K, but are ten times more frequent \citep{Teriaca:2004wy}. However, \cite{Barbieri2024b} have shown that the precise choice of the distribution $\gamma(T)$ is not very relevant, since the stationary state always exhibits temperature inversion, regardless of the choice of the distribution of the temperature increments. The choice of the distribution $\eta(t_w)$ of the waiting times between temperature increments is even less relevant, since it enters Eq.\ \eqref{eq:falphastationary} only through the constant $A$ defined in Eq.\ \eqref{eq:A}, which in turn depends only on $\langle t_w \rangle_{\eta}$, i.e., on the average value of $t_w$. 

Once the functional form of the probability distribution $\gamma$ is fixed, the distribution functions, Eq.~\eqref{eq:falphastationary}, from which all the physical quantities in the stationary state can be derived only depend on three parameters: the surface gravity $g$, the average of the temperature increments $\langle \Delta T\rangle$ and the fraction of time $A$ the thermostat spends at a temperature larger than the reference one. Choosing the following set of units, 
\begin{equation}\label{Setsofunits}
\begin{gathered}
    v_0=\sqrt{\frac{k_B T_{b}}{m_e}}, \quad m_0=m_e, \quad L_0=\frac{L}{\pi}.
\end{gathered}
\end{equation}
which imply that the unit of energy is $E_0 = k_B T_b$, where $T_b$ is the unit of temperature, the two parameters $g$ and $\langle \Delta T\rangle$ can be replaced by their dimensionless counterparts $\tilde{g}$ and $\Delta \tilde{T}$. Therefore, the three dimensionless parameters upon which the model depends are the constant $A$ given by Eq.\ \eqref{eq:A}, the strength $\tilde{g}$ of the external field in units of thermal energy,
\begin{equation}\label{eq:tilde_g}
    \tilde{g}=\frac{g(m_e+m_p)L}{2\pi k_B T_{b}}~,
\end{equation}
and the average amplitude of temperature increments $\Delta \tilde{T}$ measured in units of the reference temperature $T_b$ of the thermostat, 
\begin{equation}\label{eq:Delta_tilde_T}
\Delta \tilde{T} =\frac{\langle \Delta T\rangle}{T_b}~.
\end{equation}
The larger the parameter $\tilde{g}$ defined in Eq.\ \eqref{eq:tilde_g} above, the stronger the stratification of the atmosphere is. Therefore we shall refer to $\tilde{g}$ as the stratification parameter. 
If $A \not = 0$, the temperature $T(z)$ at any height $z$ within the loop will be strictly larger than the reference thermostat temperature $T_b$. Requiring that the temperature at the base of the loop,  $T(z=0)$, is fixed and close to the average observed value at the base of the corona, one gets an implicit relation between $A$ and $\Delta \tilde{T}$, so that the latter two parameters are no longer independent and the free parameters of the model are reduced to two, which can be chosen as $\tilde{g}$ and $\Delta \tilde{T}$. In the following we shall assume, as in \citep{Barbieri2023temperature,Barbieri2024b}, that the temperature at the base of the corona is 10\% larger than $T_b$, i.e., $T(z=0) = 1.1 \, T_b$.

\subsection{Transition region and corona}
\label{subsec:transition_region}

In Sec.~\ref{sec:main_sequence_stars_LMS} and in Appendix~\ref{sec:main_sequence_stars_HMS} we shall apply the above-described model to main-sequence stars, asking whether and under which conditions it predicts the presence of a Sun-like corona in such stars. 
To do so, we shall first examine the temperature and density profiles in two examples and clarify what does it mean, in the framework of this model, to exhibit a Sun-like corona.
Then we shall introduce an ad-hoc quantity, $X$, that depends on the parameters of the model, $\tilde{g}$ and $\Delta\tilde{T}$, and whose value is used to discriminate between the presence or absence of a Sun-like corona without having to inspect temperature and density profiles.

\begin{figure}
    \centering
    \includegraphics[width=0.99\linewidth]{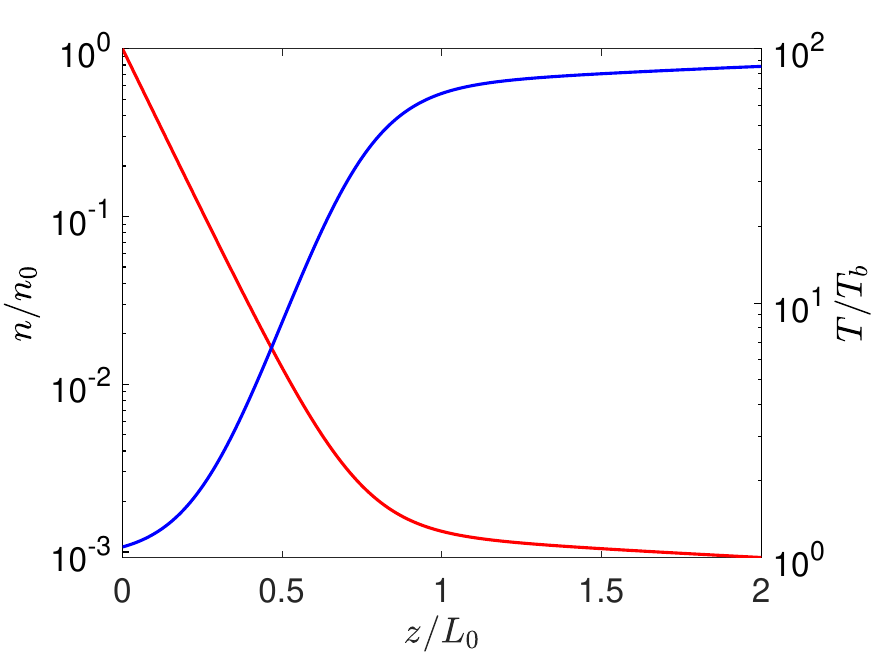}
    \caption{Temperature and density profiles as a function of the dimensionless height $z/L_0$ within the loop obtained using $\tilde{g} = 9$ and $\Delta \tilde{T} = 90$. Temperature (blue curve) is plotted in units of $T_b$, while density (red curve) is plotted in units of the density at the base of the loop, $n_0 = n(z=0)$. This is an example of Sun-like coronal profiles: for small values of $z$ there is a steep rise in $T$ (resp.\ steep fall in $n$), i.e., a transition region, at the end of which the values of $n$ and $T$ are already coronal ones, while for larger $z$'s $T$ increases (resp.\ $n$ decreases) in a much gentler way.}
    \label{fig:corona_yes}
\end{figure}

The temperature $T$ and the number density $n$ can be computed at any point of the loop from the stationary distribution functions \eqref{eq:falphastationary} according to the standard kinetic definitions \citep[see e.g.][]{nicholson1983introduction}, and turn out to be equal for both species. 
We say that the model predicts a Sun-like corona when the variation of temperature $T$ and the number density $n$ as a function of the height $z$ within the loop has the following representative properties, as shown in Fig.\ \ref{fig:corona_yes}.
First, there is a steep rise in $T$ and a steep fall in $n$ at small heights, $z$, i.e. a transition region forms.
Second, at the top of the transition region the density and temperature have already coronal values, i.e., $T \approx 10^2\, T_b$ and $n \approx 10^{-3}\, n_0$. where $n_0 = n(z = 0)$ is the density at the base of the loop.
Third, at larger heights, both the temperature increase and the density decrease are much gentler.
 
\begin{figure}
    \centering
    \includegraphics[width=0.99\linewidth]{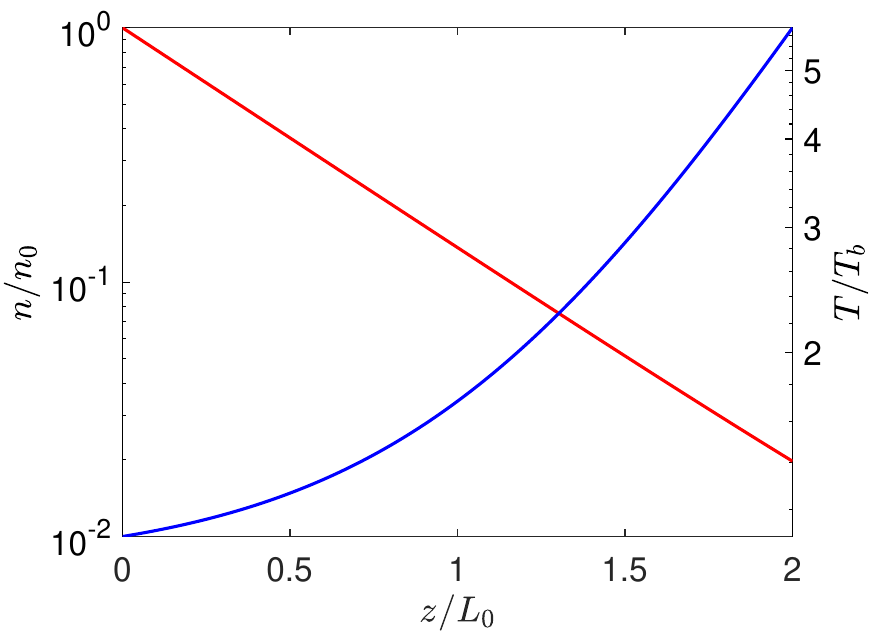}
    \caption{Temperature and density profiles as in Fig.\ \ref{fig:corona_yes}, obtained using $\tilde{g} = 2$ and $\Delta \tilde{T} = 90$. This is an example of a situation in which the model does not predict a Sun-like corona: there is temperature inversion, but the temperature reached at the top of the loop is much smaller than typical coronal temperatures and there is no transition region, since both density and temperature gently vary with $z$.}
    \label{fig:corona_no}
\end{figure}

As discussed by \citet{Barbieri2023temperature,Barbieri2024b}, these profiles are very similar to those of the atmosphere of the Sun, but for the fact that the model predicts a transition region which is thicker than the observed one. The profiles plotted in Fig.\ \ref{fig:corona_yes} have been obtained using $\tilde{g} = 9$ and $\Delta \tilde{T} = 90$.
On the contrary, the model does not predict a Sun-like corona when the profiles are as in Fig.\ \ref{fig:corona_no}, obtained using $\tilde{g} = 2$ and $\Delta \tilde{T} = 90$. Here there is no transition region, because the gradient of both $n$ and $T$ does not change very much with $z$, and the temperature reached at the top of the loop is much smaller than coronal temperature, being $T(z/L_0 = 2) \approx 5\, T_b$.  

\begin{figure}
    \centering
    \includegraphics[width=0.99\linewidth]{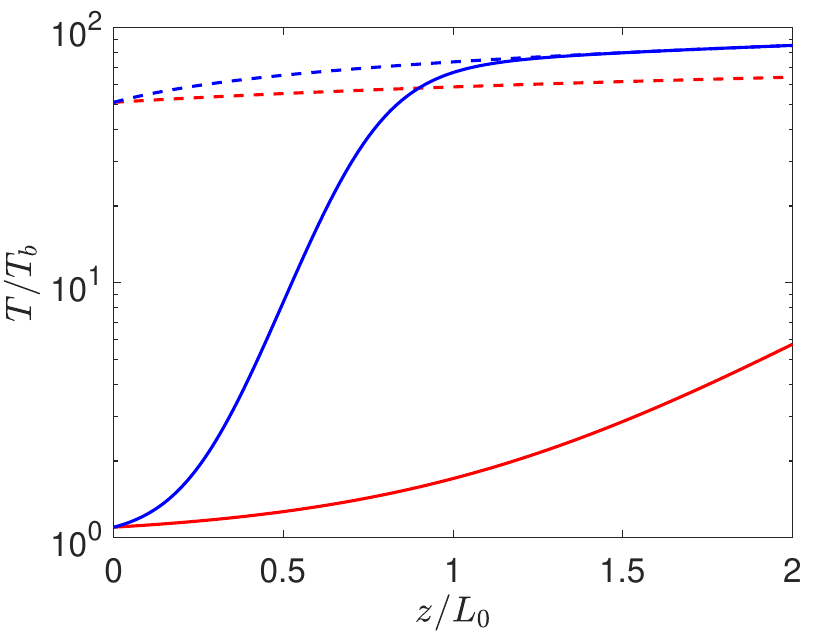}
    \caption{Comparison between the temperature profiles already shown in Figs.\ \ref{fig:corona_yes} and \ref{fig:corona_no} (solid curves) and the temperature profiles obtained using the same parameters but setting $A = 1$ (dashed curves). Blue curves refer to the case of a Sun-like corona and red curves to the case in which there is not a Sun-like corona.}
    \label{fig:A_1_A_not_1}
\end{figure}

Is it possible to define a numerical quantity whose value can discriminate between the two cases shown in Fig.\ \ref{fig:corona_yes} and \ref{fig:corona_no}, respectively? The answer is yes, as we are going to show in the following. To define such a quantity, we observe that when we do have a Sun-like corona (Fig.\ \ref{fig:corona_yes}), the large-$z$ part of the temperature profile, where $T$ does not vary much with $z$ and is close to its largest value, is almost completely coming from the rightmost contribution to the distributions function in Eq.\ \eqref{eq:falphastationary}, i.e., the integral multiplied by $A$. Indeed, by setting $A = 1$ in Eq.\ \eqref{eq:falphastationary} without changing any other parameter (which amounts to select only the second term in Eq.\ \eqref{eq:falphastationary}), and computing the temperature as before, one gets a profile which is essentially overlapping with the rightmost part of the actual temperature profile (compare the solid and dashed blue lines in Fig.\ \ref{fig:A_1_A_not_1}). 

This happens because velocity filtration allows only hot particles originating from the stochastic temperature increments (the Eq. \eqref{eq:falphastationary}) to reach the top of the loop.
The presence of a transition region at smaller $z$'s is a consequence of the rather sharp transition from a nearly-thermal distribution dominated by the contribution of the ``cold'' particles at temperature $T_b$, that is, the leftmost term in Eq.\ \eqref{eq:falphastationary}, to the strongly non-thermal distribution of the corona. On the contrary, when the values of the parameters are such that the model does not describe a Sun-like corona, as in Fig.\ \ref{fig:corona_yes}, the velocity filtration mechanism is not so efficient and the contribution from the ``hot'' particles never dominates the distribution function. By computing the temperature setting $A = 1$ in this case one obtains a curve whose values are always much larger than the actual temperatures (compare the solid and dashed lines in Fig.\ \ref{fig:A_1_A_not_1}).    
Therefore, we define the following quantity:
\begin{equation}\label{parameterX}
X(\tilde{g},\Delta{\tilde{T}})=\frac{T(z=z_t)}{T(z=z_t,A=1)}~,
\end{equation}
where $T(z=z_t)$ is the temperature at the top height $z = z_t$ of the loop (i.e., at $z/L_0 = 2$) while $T(z=z_t,A=1)$ is the temperature at $z = z_t$ computed by keeping the same values of $\tilde{g}$ and $\Delta{\tilde{T}}$ but setting $A = 1$ in Eq.\ \eqref{eq:falphastationary}. The above discussion implies that when $X \approx 1$ the model predicts a Sun-like corona, as in the case depicted in Fig.\ \ref{fig:corona_yes}, while if $X$ is definitely smaller than one we are in the situation of Fig.\ \ref{fig:corona_no} and the model does not predict a Sun-like corona. In particular, Fig.\ \ref{fig:corona_yes} corresponds to $X = 1$ (up to $2\times 10^{-5}$) and  Fig.\ \ref{fig:corona_no} corresponds to $X \simeq 0.09$. For practical purposes, we set a threshold at $X = 0.9$ and say that when $X>0.9$ we have a Sun-like corona, while if $X<0.9$ we do not.

\section{Application to low-mass main-sequence stars}
\label{sec:main_sequence_stars_LMS}
Let us now apply our model to low-mass main-sequence stars, i.e., stars with $M<1.5\, M_{\odot}$, for which a Sun-like corona is expected to be present (in  Appendix \ref{sec:main_sequence_stars_HMS} we will discuss the case of larger-mass stars).

In the previous sections we have shown that the predictions of the model only depend on two dimensionless parameters, $\tilde{g}$ and $\Delta \tilde{T}$. Both dimensionless parameters are fixed by three star quantities, i.e., the star mass $M$, the star radius $R$ and the thermostat reference temperature $T_b$ (which we identify with the typical temperature of the star's high chromosphere), and by two quantities defining the model, i.e., the loop length $L$ and the average of temperature increments $\langle \Delta T \rangle$. Using the scaling relations valid for low-mass main sequence stars \citep{Eker2018} we can express $\tilde{g}$ and $\Delta{\tilde{T}}$ in terms of $M$, $L$ and $\langle \Delta T \rangle$. To do so, we first use the scaling law relating the star's radius to its mass, namely
\begin{equation}\label{Scalingradius}
     \frac{R}{R_{\odot}}=a\left(\frac{M}{M_{\odot}}\right)^{2}+b\left(\frac{M}{M_{\odot}}\right)+c~,
\end{equation}
where $R_{\odot}$ and $M_{\odot}$ are the Sun's radius and mass, respectively, $a=0.438$, $b=0.479$, and $c=0.075$. We also use the scaling law that relates the surface temperature $T$ of a star to its mass, i.e.,
\begin{equation}\label{Scalingtempbase_surf}
     \frac{T}{T_{\odot}}=\left[\frac{\mathcal{L}}{\mathcal{L}_{\odot}}\left(\frac{R}{R_{\odot}}\right)^{-2}\right]^{1/4}~, 
\end{equation}
where $T_{\odot}$ is the surface temperature of the Sun,
\begin{equation}\label{Scalingtempbase_surf_2}
     \frac{\mathcal{L}}{\mathcal{L}_{\odot}} = \left(\frac{M}{M_{\odot}}\right)^{\alpha}10^d\,,    
\end{equation}
and the parameters $\alpha$ and $d$ in Eq.\ \eqref{Scalingtempbase_surf_2} are given by
\begin{equation} \label{eq:scalingparameter_a}
    \begin{split}
        \alpha & = 2.028 \quad d= -0.976 \quad \text{if} ~ \quad 0.179 < \frac{M}{M_{\odot}} < 0.45 \,; \\
        \alpha & = 4.572 \quad d= -0.102 \quad \text{if} ~ \quad 0.45 < \frac{M}{M_{\odot}} < 0.72 \,; \\
        \alpha & = 5.743 \quad d= -0.007 \quad \text{if} ~ \quad 0.72 < \frac{M}{M_{\odot}} < 1.05 \,; \\
        \alpha & = 4.329 \quad d= 0.010 \quad \text{if} ~ \quad 1.05 < \frac{M}{M_{\odot}} < 1.5 \ . \\
    \end{split}
\end{equation}

\begin{figure}
    \centering
    \includegraphics[width=0.99\columnwidth]{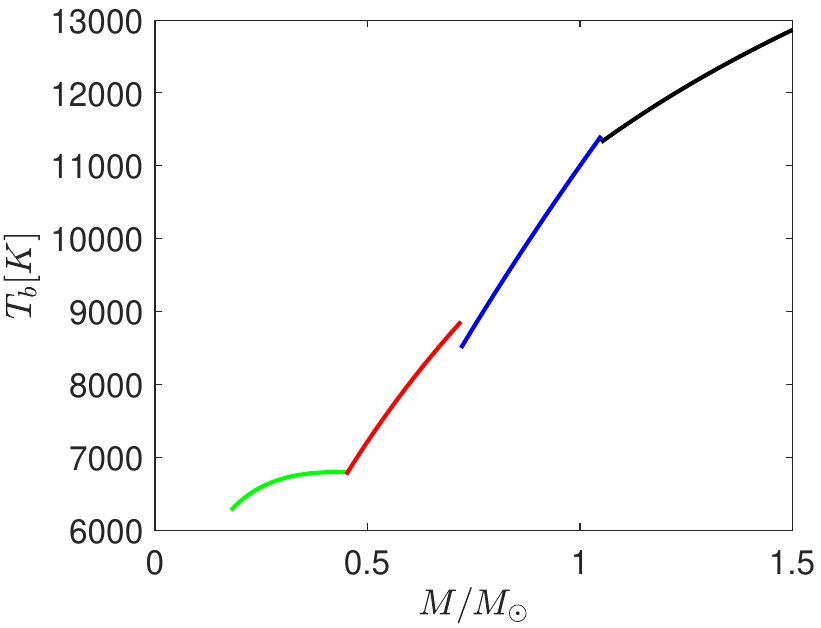}
    \caption{
    Plot of the reference temperature of the thermostat (chromosphere), $T_b$ in Kelvin, as a function of the mass $M$ (in units of solar mass $M_{\odot}$). The green curve corresponds to the mass interval $0.179 < M/M_{\odot} < 0.45$, the red curve to $0.45 < M/M_{\odot} < 0.72$, the blue curve to $0.72 < M/M_{\odot} < 1.05$ and finally the black one to $1.05 < M/M_{\odot} < 1.5$.}
    \label{fig:Tbase}
\end{figure}

Let us now assume that Eq.\ \eqref{Scalingtempbase_surf}, valid for surface temperatures, also holds for the chromospheric temperatures at the base of the loop which we take as thermostat reference temperatures in the model, i.e., $T_b$, so that we can write 
\begin{equation}\label{Scalingtempbase}
     \frac{T_b}{T_{b,\odot}}= \left[\frac{\mathcal{L}}{\mathcal{L}_{\odot}}\left(\frac{R}{R_{\odot}}\right)^{-2}\right]^{1/4}~, 
\end{equation}
where $T_{b,\odot}$ is the reference thermostat temperature for the Sun, i.e., $T_{b,\odot} = 10^4$ K, and $\mathcal{L}/\mathcal{L}_\odot$ is given by Eq.\ \eqref{Scalingtempbase_surf_2}. In Fig.\ \ref{fig:Tbase}, $T_b$ is plotted as a function of $M/M_{\odot}$. Using Eqs.\ \eqref{Scalingradius}, \eqref{Scalingtempbase_surf_2} and \eqref{Scalingtempbase} we can write the dimensionless parameter $\Delta \tilde{T}$ for a generic star as a function of the mass $M$ of the star:
\begin{equation}
\label{eq:DeltaTstarLMS}
    \Delta \tilde{T} = \frac{\langle \Delta T \rangle}{T_{b,\odot}} 10^{-d/4} \left(\frac{M}{M_{\odot}}\right)^{-\alpha/4} \left[a\left(\frac{M}{M_{\odot}}\right)^{2}+b\left(\frac{M}{M_{\odot}}\right)+c\right]^{1/2} ~.
\end{equation}

\begin{figure}
    \centering
    \includegraphics[width=0.99\columnwidth]{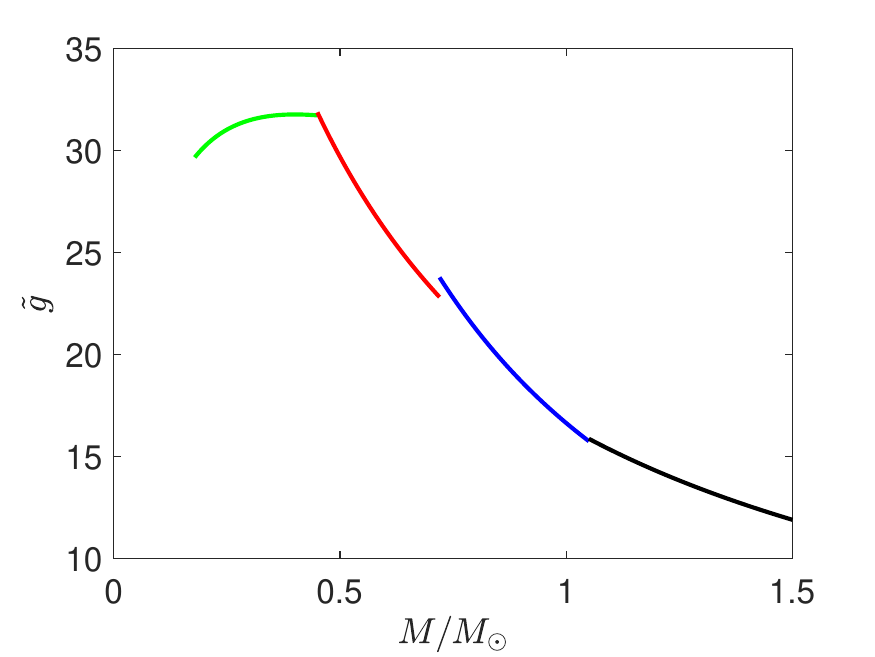}
    \caption{As in Fig. \ref{fig:Tbase}, for the stratification parameter $\tilde{g}$ as given by Eq.\ \eqref{parametergstars} with $z_t/R = 1/35$.}
    \label{fig:gtilde}
\end{figure}

From now on, we fix the value of the average temperature increment in Eq.\ \eqref{eq:DeltaTstarLMS} to the one we used to produce the plots on Figs.\ \ref{fig:corona_yes}, \ref{fig:corona_no} and \ref{fig:A_1_A_not_1}, that is, $\langle \Delta T \rangle = 9 \times 10^5$ K. The reason for this choice is that with such a value we have a million-Kelvin corona in the case of the Sun, as shown by \citet{Barbieri2023temperature,Barbieri2024b}, and if we assume that the mechanism producing temperature fluctuations in the chromosphere of a generic main-sequence star is the same at work on the Sun, it is natural to assume that it produces the same temperature increments. We shall nonetheless discuss the effects of varying $\langle \Delta T \rangle$ later, in Sec.\ \ref{sec:amplitude_temp_fluc}. Finally, using Eqs.\ \eqref{Scalingradius}  and \eqref{Scalingtempbase} we express the stratification parameter $\tilde{g}$ in terms of the mass $M$ of the star and of the loop height $z_{t}$ as 
\begin{equation}\label{parametergstars}
\tilde{g}=\alpha_{\odot} \frac{z_{t}}{R} \mathcal{B}\left(\frac{M}{M_{\odot}}\right)~, 
\end{equation}
where the function $\mathcal{B}$ is
\begin{equation}\label{parametergstars_2}
\mathcal{B}\left(\frac{M}{M_{\odot}}\right) = 10^{-d/4} \left(\frac{M}{M_{\odot}}\right)^{1 - \frac{\alpha}{4}} \left[a\left(\frac{M}{M_{\odot}}\right)^{2}+b\left(\frac{M}{M_{\odot}}\right)+c\right]^{-1/2}
\end{equation}

and the constant $\alpha_\odot$ is given by
\begin{equation}\label{parameteralphasun}
    \alpha_{\odot} = \frac{G M_{\odot}}{R_{\odot}}\frac{m_e+m_p}{4k_B T_{b,\odot}}~.
\end{equation}
A plot of $\tilde{g}$ as a function of $M/M_{\odot}$ for a fixed value of $z_t$ is shown in Fig.\ \ref{fig:gtilde}.

\subsection{Presence of a Sun-like corona}
\label{subsec:sunlike_coronaLMS}
\begin{figure}
    \centering
    \includegraphics[width=0.99\columnwidth]{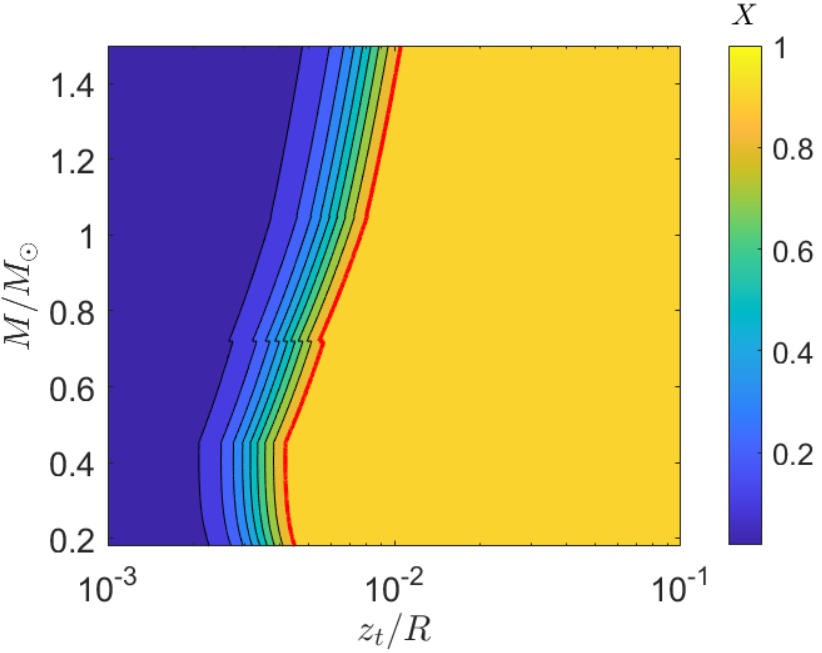}
    \caption{Contour plot of $X$ as defined in Eq.\ \eqref{parameterX}, computed for low-mass stars (i.e., $M \in [0.179M_{\odot},1.5\, M_{\odot}]$). $X$ is plotted as a function of the star mass $M$ in units of solar mass $M_{\odot}$ and of the top height of the loop $z_t$ scaled by the star radius $R$. The red line corresponds to $X=0.9$, the threshold separating the regime without a Sun-like corona ($X < 0.9$, bluish colours) from that where there is a Sun-like corona ($X > 0.9$, yellowish colours).}
    \label{fig:ContoursmallM}
\end{figure}

To understand whether the model predicts a Sun-like corona for low-mass main-sequence stars, in Fig. \ref{fig:ContoursmallM} we plot the values of the quantity $X$ defined in Eq.\ \eqref{parameterX} as a function of $z_{t}/R$ and $M/M_{\odot}$. In defining the model we made the approximation of constant gravity, i.e., of a gravity force equal to the value at the surface of the star throughout the loop. In order for this approximation to be reasonable, we chose to consider values of $z_t$ such that $z_{t}/R \le 0.1$. The transition between the two regimes, with and without a Sun-like hot corona, is marked by the red line corresponding to the threshold value $X = 0.9$ in figure \ref{fig:ContoursmallM}.

Figure \ref{fig:ContoursmallM} show that, according to the model, all low-mass main-sequence stars have a million-degree, Sun-like corona, because coronal conditions are met whenever one gets sufficiently high in the star's atmosphere. For solar masses, $z_{t}/R > 0.01$ is sufficient to have a million-degree corona, and this minimum height gets smaller for smaller masses, reaching $z_{t}/R > 0.0045$ when $M = 0.179\, M_\odot$. 
 
Inspection of figure \ref{fig:ContoursmallM} shows that it is ``easier'' to have a Sun-like corona as the mass decreases, until $M = 0.5 M_{\odot}$; for smaller masses, this trend reverses. This is a consequence of the fact that larger values of $\tilde{g}$ favour the presence of a Sun-like corona and indeed $\tilde{g}$ rapidly increases passing from $1.5 M_{\odot}$ to $0.5 M_{\odot}$, and then starts decreasing for smaller masses, as can be seen in figure \ref{fig:gtilde}.
In all the mass regime the minimum height to reach coronal conditions (i.e., such that $X>0.9$) is always much smaller than $0.1\, R$, therefore the approximation of constant gravity is well justified. 

\begin{figure}
    \centering
    \includegraphics[width=0.99\columnwidth]{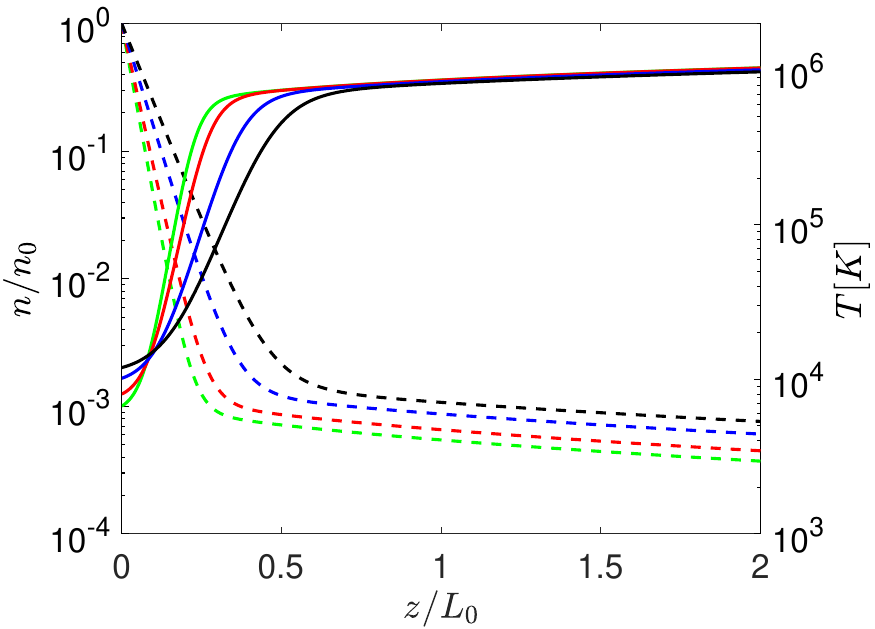}
    \caption{Density $n$ (in units of the density at the base of the loop, $n_0 = n(z = 0)$) and temperature $T$ (in Kelvin) as a function of the height $z$ within the loop, scaled by $L_0 = L/\pi$ (where $2L$ is the loop length), for some values of the mass in the range $0.179 < M/M_{\odot} < 1.5$. Here we choose $z_t/R = 1/35$, such that $X > 0.9$ for all the values of $M$ and there always is a Sun-like corona. Solid lines correspond to temperatures and dashed lines to densities. The green curves are computed for a star with mass $M = 0.3 \, M_{\odot}$, the red ones for $M = 0.6 \, M_{\odot}$, the blue ones for $M = 0.8 \, M_{\odot}$ and the black ones for $M = 1.2 \, M_{\odot}$.
    }
    \label{fig:TvsnMsmallMsun}
\end{figure}

In order to show how the density and temperature profiles depend on $M$, we set $z_{t}/R=1/35$, i.e., a value such that $X > 0.9$ for any value of $M$, and we plot in figure \ref{fig:TvsnMsmallMsun} the temperature $T$ (in Kelvin) 
and the density $n$ (in units of the density $n_0$ at the base of the loop) as a function of the height $z$ within the loop (in units of $L_0 = L/\pi$, as in Figs.\ \ref{fig:corona_yes} and \ref{fig:corona_no}). 
Figure \ref{fig:TvsnMsmallMsun} shows that although the temperature at the base of the loop increases with the mass $M$ of the star (as implied by Eqs.\ \eqref{Scalingtempbase_surf} and \eqref{Scalingtempbase} and shown in Fig.\ \ref{fig:Tbase}) the temperature at the top is about one million degrees for all the values of $M$; this is due to the fact that the ``cold'' particles are gravitationally filtered out in the corona, leaving only the ``hot'' particles generated by the temperature increments at $\Delta T = 9\times10^5~\mathrm{K}$. Moreover, figure \ref{fig:TvsnMsmallMsun} shows that for low-mass stars the transition region becomes steeper and steeper as the mass of star $M$ decreases. This can be understood by looking at Fig.\ \ref{fig:ContoursmallM}. The red level curve corresponding to the threshold $X=0.9$ moves towards smaller values of $z_t$ as the mass of the star decreases and the level curves of $X$ become denser. As a consequence, $\tilde{g}$ increases as $M$ decreases, as we already noted above. Equation \eqref{eq:falphastationary} in turn implies that as $\tilde{g}$ increases, the population of ``cold'' particles thermally distributed at the mean chromospheric temperature $T_b$ is more and more exponentially depressed with height $z$. Therefore, by increasing the value of $\tilde{g}$ we expect an increasingly steep transition region.
 The density profile is anti-correlated with that of the temperature for all the values of $M$. The coronal density (in units of the density at the base of the loop, $n_0$) decreases with $M$, again because $\tilde{g}$ increases: given the density at the base, by increasing the value of $\tilde{g}$ fewer and fewer particles can reach the top of the loop, and the coronal density decreases. This notwithstanding, the density drop in the corona with respect to the density at the base of the loop is never much larger than that on the Sun, being between three and four orders of magnitudes. 

\subsection{Amplitude of temperature fluctuations}
\label{sec:amplitude_temp_fluc}
\begin{figure}
    \centering
    \includegraphics[width=0.99\columnwidth]{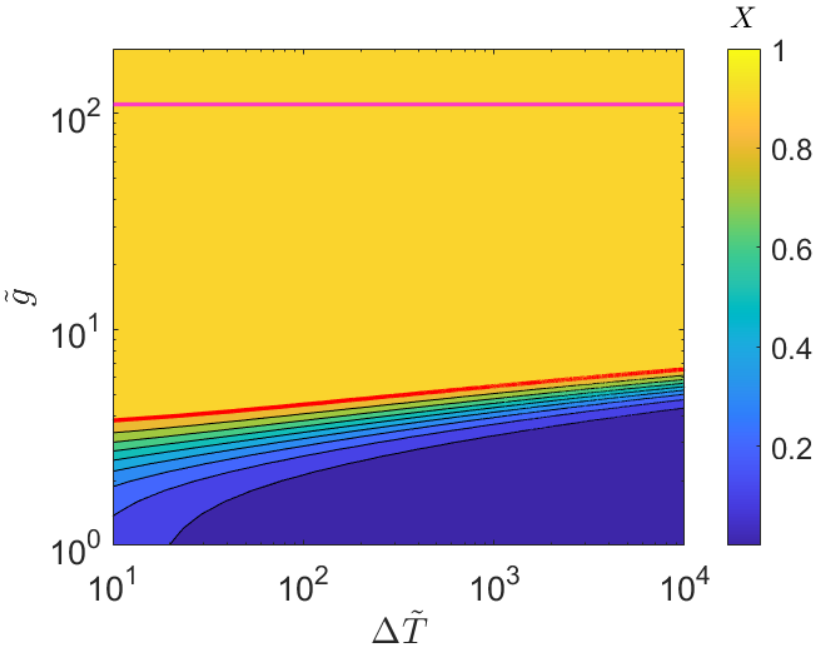}
    \caption{
    Contour plot of $X$ computed using Eq.\ \eqref{parameterX} as a function of the dimensionless intensity of the temperature increments $\Delta {\tilde{T}}$ and of the stratification parameter $\tilde{g}$. The red solid line corresponds to the threshold value $X = 0.9$ separating the region of parameters in which the model predicts a Sun-like transition region and corona ($X > 0.9$, yellowish colours) from that where it does not ($X < 0.9$, bluish colours). The magenta horizontal line 
    is the maximum value of $\tilde{g}$ as a function of $M$.}
    \label{fig:ContourTpg}
\end{figure}

Throughout our discussion we have set the parameter $\langle \Delta T \rangle = \langle \Delta T_{\odot} \rangle = 9\times 10^5$ K based on the assumption that the physical processes responsible for stochastic temperature increments in the high chromosphere of a generic main-sequence star are the same acting on the Sun and producing a million-Kelvin corona \citep[see][]{Barbieri2023temperature,Barbieri2024b}. In our model, the value of $\langle \Delta T \rangle$ is crucial to determine the coronal temperature: choosing, say, values an order of magnitude larger or smaller would shift the resulting coronal temperature accordingly. Therefore, if we think that main-sequence stars' coronae have temperatures similar to the Sun's corona, the choice we made for $\langle \Delta T \rangle$ seems appropriate. This notwithstanding, one may wonder whether the values of the average temperature fluctuations only settle the coronal temperature or influence also other properties of the coronal temperature and density profiles. To answer this question, in Fig.\ \ref{fig:ContourTpg} we plot $X$ as a function of $\tilde{g}$ and $\Delta \tilde{T}$; the latter is the dimensionless parameter directly related to $\langle \Delta T \rangle$, see Eq.\ \ref{eq:DeltaTstarLMS}. We consider over three decades of $\Delta{\tilde{T}}$, showing that the specific value of $\Delta{\tilde{T}}$ does not strongly affect the transition between the region where $X<0.9$ (so that no Sun-like transition region is present) and the region where $X > 0.9$ (where we have a Sun-like transition region and a corona). The threshold between the two regions, marked by the red curve in Fig.\ \ref{fig:ContourTpg}, stays almost constant as a function of $\Delta{\tilde{T}}$ in a range of $\tilde{g}$ values that includes the maximum value of $\tilde{g}$ for low-mass stars. Maximum of $\tilde{g}$ has been evaluated from Eq.\ \eqref{parametergstars} within the range of $M$ and $z_t$ considered above. 
Therefore, the specific value of $\Delta \tilde{T}$ does not have a strong impact on the properties of the temperature profiles (e.g., on the size and shape of the transition region).

\subsection{Relaxation and fluctuation timescales}
\label{sec:timescales}
The model is valid only if the timescales of the fluctuating thermostat are fast enough to prevent the coronal plasma from relaxing towards a thermal equilibrium state, i.e., if conditions \eqref{eq:time_constraints} are satisfied. However, these timescales depend on the mass of the star. For a given main-sequence star of mass $M$, the relaxation time $t_r$ 
is still defined by Eq.\  \eqref{relaxationtimes}, with $t_T$ now given by
\begin{equation}
    t_T = \frac{z_{t}}{z_{t,\odot}} \sqrt{\frac{\langle \Delta T_{\odot} \rangle}{\langle \Delta T \rangle}} t_{T_{\odot}} \quad \text{with} \quad t_{T_{\odot}}=2L_{\odot} \sqrt{\frac{m_e}{k_B \langle \Delta T_{\odot}\rangle }}~,
\end{equation}
and $t_{g}$ by
\begin{equation}
    t_{g} = \sqrt{\frac{z_{t}}{z_{t,\odot}}} \sqrt{\frac{M_{\odot}}{M}}\frac{R}{R_{\odot}} t_{g_{\odot}} \quad \text{with} ~ t_{g_{\odot}}=\sqrt{\frac{2 L_{\odot} m_e}{g_{\odot}(m_e+m_p)}}~,
\end{equation}
where $t_{T,\odot}$ and $t_{g,\odot}$ are the thermal and gravitational crossing times for the case of the Sun and we have assumed, as above, $\langle \Delta T \rangle = \langle \Delta T_\odot \rangle$. Using $L_{\odot}=\pi \times 10^9$ cm (typical for coronal loops in the Sun's atmosphere) and $\langle \Delta T_{\odot} \rangle = 9 \times 10^5$ K, as done by \citet{Barbieri2023temperature}, to calculate the relaxation time scale for the Sun as $t_{r_{\odot}}=\mathrm{min}[t_{T_{\odot}}, t_{g_{\odot}}]$ we obtain $t_{r_{\odot}} \approx 10$ s. For low-mass main-sequence stars, considering the same range of values of $z_t$ as in Fig.\ \ref{fig:ContoursmallM}, we get
$0.1\, t_{r_{\odot}} \le t_{r} \le 7\, t_{r_{\odot}}$. Therefore, we obtain values of the relaxation time between 1 s and 70 s, i.e., within two orders of magnitude from those of the Sun for all low-mass main-sequence stars.

\section{Conclusions and perspectives}
\label{sec:conclusions}

We have investigated the role of stochastic temperature fluctuations in the high chromosphere in shaping the temperature and density profiles of the coronae of main-sequence stars. We have used a model of a plasma atmosphere confined by the magnetic field (i.e., made up of coronal loops) in thermal contact with the chromosphere, the latter being modelled as a thermostat with fluctuating temperature. This model was recently put forward by \citet{Barbieri2023temperature,Barbieri2024b} and has already been applied to the solar atmosphere, successfully reproducing its inverted density and temperature profiles. In this model, stochastic temperature fluctuations at the base of the loop structures produce a non-thermal population of ``hot'' and fast particles that can climb the gravity well and form the corona. On the contrary, ``cold'' particles that are thermally distributed at the mean chromospheric temperature mostly stay close to the base of the loop structures. The transition region is where the two populations coexist, with a relative abundance strongly depending on the height above the surface of the star. We have applied the above formalism to the case of low-mass main-sequence stars, where we expect that a corona may form out of stationary loops and therefore our model should be applicable (the case of larger-mass main-sequence stars is addressed in Appendix \ref{sec:main_sequence_stars_HMS}). We have shown that the model always predicts inverted temperature and density profiles with a transition region and a corona for all main-sequence stars. Furthermore, the model predicts a transition region that becomes steeper as the mass of the star $M$ decreases. 

Concerning possible future developments, we first note that, according to our model, the size of the transition region relative to the total size of the system is determined by the stratification parameter $\tilde{g}$. This is due to the fact that the plasma in the loop is treated as collisionless, 
so the stronger the gravitational energy is in units of thermal energy at the bottom, the less the conservation of single-particle energy allows cold particles to climb the potential well, and consequently the narrower the transition region. Therefore, being $\tilde{g}$ a function of $M$, the fact that the size of the transition region depends on the stellar mass in our model is entirely due to this effect. However, although not-so-frequent, collisions between particles do occur in the solar corona \citep{Aschwandensolarcorona}. As shown by \citet{Landi-Pantellini2001}, collisions are another ingredient that can determine the shape of the transition region. Since the mean free path of a particle in a plasma scales with velocity as $v^4$, where $v$ is the particle velocity, we still expect that ``cold'' particles would populate the base of the atmosphere; the ``hot'' ones would reach coronal heights at a distance from the base governed by the mean free path. Therefore, in order to fully understand how temperature fluctuations at the base of a stellar atmosphere may shape a transition region and produce a million-Kelvin corona, it would be interesting to study the interplay between the role of gravitational filtering (studied in the present work) and the filtering effect produced by Coulomb collisions. 

As a further possible development, we note that the atmospheres of stars are not static, but evolve in the form of stellar winds \citep{Parker1958}. For this reason, an extension of our modelling to the case of open geometry could be interesting. The exospheric approach \citep{Chamberlain1960,Jockers1970,Lemaire1971,Lamy2003,Maksimovic1997,Zouganelis2004}, which explains the formation of a stellar wind as a collisionless evaporation from a given altitude, seems the relevant one for the extension of our work: indeed, using the same formalism of the present work, it might be possible to build an exospheric model having its base in the high chromosphere and able to reproduce not only the plasma of the transition region and of the corona but also the stellar wind.
 
\begin{acknowledgements}
We wish to thank Germano Sacco and Elena Pancino for a careful reading of the manuscript and for very useful comments that helped us to improve our work. We acknowledge partial financial support from the Solar Orbiter/Metis program supported by the Italian Space Agency (ASI) under the contracts to the National Institute of Astrophysics (INAF), Agreement ASI-INAF N.2018-30-HH.0.
This research was partially funded by the European Union - Next Generation EU - National Recovery and Resilience Plan (NRRP) - M4C2 Investment 1.4 - Research Programme  CN00000013 "National Centre for HPC, Big Data and Quantum Computing" - CUP B83C22002830001 and by the European Union - Next Generation EU - National Recovery and Resilience Plan (NRRP)- M4C2 Investment 1.1- PRIN 2022 (D.D. 104 del 2/2/2022) - Project `` Modeling Interplanetary Coronal Mass Ejections'', MUR code 31. 2022M5TKR2,  CUP B53D23004860006.
Views and opinions expressed are however those of the author(s) only and do not necessarily reflect those of the European Union or the European Commission. Neither the European Union nor the European Commission can be held responsible for them.
\end{acknowledgements}

   \bibliographystyle{aa} 
   \bibliography{manuscript} 

\providecommand{\noopsort}[1]{}\providecommand{\singleletter}[1]{#1}%
\begin{thebibliography}{45}
\expandafter\ifx\csname natexlab\endcsname\relax\def\natexlab#1{#1}\fi

\bibitem[{{Aschwanden}(2005)}]{Aschwandensolarcorona}
{Aschwanden}, M.~J. 2005, {Physics of the Solar Corona. An Introduction with
  Problems and Solutions} (Springer), 2nd edition

\bibitem[{Barbieri {et~al.}(2024{\natexlab{a}})Barbieri, Casetti, Verdini, \&
  Landi}]{Barbieri2023temperature}
Barbieri, L., Casetti, L., Verdini, A., \& Landi, S. 2024{\natexlab{a}}, A\&A,
  681, L5

\bibitem[{Barbieri {et~al.}(2024{\natexlab{b}})Barbieri, Papini, Di~Cintio,
  Landi, Verdini, \& Casetti}]{Barbieri2024b}
Barbieri, L., Papini, E., Di~Cintio, P., {et~al.} 2024{\natexlab{b}}, Journal
  of Plasma Physics, 90, 905900511

\bibitem[{Belmont {et~al.}(2013)Belmont, Grappin, Mottez, Pantellini, \&
  Pelletier}]{belmont2013collisionless}
Belmont, G., Grappin, R., Mottez, F., Pantellini, F., \& Pelletier, G. 2013,
  Collisionless Plasmas in Astrophysics (Wiley)

\bibitem[{{Berghmans} {et~al.}(2021){Berghmans}, {Auch{\`e}re}, {Long},
  {Soubri{\'e}}, {Mierla}, {Zhukov}, {Sch{\"u}hle}, {Antolin}, {Harra},
  {Parenti}, {Podladchikova}, {Aznar Cuadrado}, {Buchlin}, {Dolla}, {Verbeeck},
  {Gissot}, {Teriaca}, {Haberreiter}, {Katsiyannis}, {Rodriguez}, {Kraaikamp},
  {Smith}, {Stegen}, {Rochus}, {Halain}, {Jacques}, {Thompson}, \&
  {Inhester}}]{Berghmans:2021wl}
{Berghmans}, D., {Auch{\`e}re}, F., {Long}, D.~M., {et~al.} 2021, Astron.\
  Astrophys., 656, L4

\bibitem[{{Chamberlain}(1960)}]{Chamberlain1960}
{Chamberlain}, J.~W. 1960, Astrophysical Journal, 131, 47

\bibitem[{{Dere} {et~al.}(1989){Dere}, {Bartoe}, \& {Brueckner}}]{Dere:1989ux}
{Dere}, K.~P., {Bartoe}, J. D.~F., \& {Brueckner}, G.~E. 1989, Solar Physics,
  123, 41

\bibitem[{Dmitruk \& Gomez(1997)}]{Dmitruk:1997uf}
Dmitruk, P. \& Gomez, D.~O. 1997, Astrophysical Journal Letters v.484, 484, L83

\bibitem[{Eker {et~al.}(2018)Eker, Bakış, Bilir, Soydugan, Steer, Soydugan,
  Bakış, Aliçavuş, Aslan, \& Alpsoy}]{Eker2018}
Eker, Z., Bakış, V., Bilir, S., {et~al.} 2018, Monthly Notices of the Royal
  Astronomical Society, 479, 5491

\bibitem[{Fisher {et~al.}(1998)Fisher, Longcope, Metcalf, \&
  Pevtsov}]{Fisher_1998}
Fisher, G.~H., Longcope, D.~W., Metcalf, T.~R., \& Pevtsov, A.~A. 1998, The
  Astrophysical Journal, 508, 885

\bibitem[{Golub \& Pasachoff(2009)}]{GolubPasachoff:book}
Golub, L. \& Pasachoff, J.~M. 2009, The Solar Corona, 2nd edn. (Cambridge:
  Cambridge University Press)

\bibitem[{Gudel(2004)}]{Gdel2004}
Gudel, M. 2004, The Astronomy and Astrophysics Review, 12

\bibitem[{{Gudiksen} \& {Nordlund}(2005)}]{2005ApJ...618.1020G}
{Gudiksen}, B.~V. \& {Nordlund}, {\r{A}}. 2005, Astrophys.\ J., 618, 1020

\bibitem[{{Heyvaerts} \& {Priest}(1983)}]{Heyvaerts_Priest_1983}
{Heyvaerts}, J. \& {Priest}, E.~R. 1983, Astron.\ Astrophys., 117, 220

\bibitem[{{Howson} {et~al.}(2020){Howson}, {De Moortel}, \&
  {Reid}}]{2020A&A...636A..40H}
{Howson}, T.~A., {De Moortel}, I., \& {Reid}, J. 2020, Astron.\ Astrophys.,
  636, A40

\bibitem[{{Ionson}(1978)}]{Ionson_1978}
{Ionson}, J.~A. 1978, Astrophys.\ J., 226, 650

\bibitem[{{Jockers}(1970)}]{Jockers1970}
{Jockers}, K. 1970, Astron.\ Astrophys., 6, 219

\bibitem[{{Klimchuk}(2006)}]{Klimchuk_2006}
{Klimchuk}, J.~A. 2006, Solar Physics, 234, 41

\bibitem[{{Lamy} {et~al.}(2003){Lamy}, {Pierrard}, {Maksimovic}, \&
  {Lemaire}}]{Lamy2003}
{Lamy}, H., {Pierrard}, V., {Maksimovic}, M., \& {Lemaire}, J.~F. 2003, Journal
  of Geophysical Research (Space Physics), 108, 1047

\bibitem[{Landi \& Pantellini(2001)}]{Landi-Pantellini2001}
Landi, S. \& Pantellini, F. G.~E. 2001, Astron.\ Astrophys., 372, 686

\bibitem[{{Lemaire} \& {Scherer}(1971)}]{Lemaire1971}
{Lemaire}, J. \& {Scherer}, M. 1971, J.\ Geophys.\ Res., 76, 7479

\bibitem[{{Maksimovic} {et~al.}(1997){Maksimovic}, {Pierrard}, \&
  {Lemaire}}]{Maksimovic1997}
{Maksimovic}, M., {Pierrard}, V., \& {Lemaire}, J.~F. 1997, Astron.\
  Astrophys., 324, 725

\bibitem[{Maoz(2007)}]{Maoz:1142636}
Maoz, D. 2007, {Astrophysics in a nutshell; 1st ed.} (Princeton, NJ: Princeton
  Univ. Press)

\bibitem[{Molnar {et~al.}(2019)Molnar, Reardon, Chai, Gary, Uitenbroek, Cauzzi,
  \& Cranmer}]{Molnar_2019}
Molnar, M.~E., Reardon, K.~P., Chai, Y., {et~al.} 2019, Astrophys.\ J., 881, 99

\bibitem[{{Ness, J.-U.} {et~al.}(2004){Ness, J.-U.}, {Güdel, M.}, {Schmitt, J.
  H. M. M.}, {Audard, M.}, \& {Telleschi, A.}}]{Xraycoronae}
{Ness, J.-U.}, {Güdel, M.}, {Schmitt, J. H. M. M.}, {Audard, M.}, \&
  {Telleschi, A.} 2004, A\&A, 427, 667

\bibitem[{Nicholson(1983)}]{nicholson1983introduction}
Nicholson, D.~R. 1983, Introduction to Plasma Theory (Wiley)

\bibitem[{Pallavicini(1989)}]{Pallavicini1989}
Pallavicini, R. 1989, The Astronomy and Astrophysics Review, 1, 177–207

\bibitem[{{Pannekoek}(1922)}]{Pannekoek_1922}
{Pannekoek}, A. 1922, Bull. Astr. Inst. Netherlandd, 1, 107

\bibitem[{{Parker}(1958)}]{Parker1958}
{Parker}, E.~N. 1958, Astrophys.\ J., 128, 664

\bibitem[{Parker(1972)}]{Parker:1972wu}
Parker, E.~N. 1972, Astrophysical Journal, 174, 499

\bibitem[{{Parnell} \& {De Moortel}(2012)}]{2012coronalheating}
{Parnell}, C.~E. \& {De Moortel}, I. 2012, Philosophical Transactions of the
  Royal Society of London Series A, 370, 3217

\bibitem[{{Peter} {et~al.}(2014){Peter}, {Tian}, {Curdt}, {Schmit}, {Innes},
  {De Pontieu}, {Lemen}, {Title}, {Boerner}, {Hurlburt}, {Tarbell}, {Wuelser},
  {Mart{\'\i}nez-Sykora}, {Kleint}, {Golub}, {McKillop}, {Reeves}, {Saar},
  {Testa}, {Kankelborg}, {Jaeggli}, {Carlsson}, \& {Hansteen}}]{Peter:2014uz}
{Peter}, H., {Tian}, H., {Curdt}, W., {et~al.} 2014, Science, 346, 1255726

\bibitem[{Pevtsov {et~al.}(2003)Pevtsov, Fisher, Acton, Longcope, Johns-Krull,
  Kankelborg, \& Metcalf}]{Pevtsov_2003}
Pevtsov, A.~A., Fisher, G.~H., Acton, L.~W., {et~al.} 2003, The Astrophysical
  Journal, 598, 1387

\bibitem[{Pontieu {et~al.}(2011)Pontieu, Mcintosh, Carlsson, Viggo~H, Tarbell,
  Boerner, Martinez-Sykora, Schrijver, \& Title}]{Pontieu:2011vg}
Pontieu, B.~D., Mcintosh, S.~W., Carlsson, M., {et~al.} 2011, Science, 331, 55

\bibitem[{{Raouafi} {et~al.}(2023){Raouafi}, {Stenborg}, {Seaton}, {Wang},
  {Wang}, {DeForest}, {Bale}, {Drake}, {Uritsky}, {Karpen}, {DeVore},
  {Sterling}, {Horbury}, {Harra}, {Bourouaine}, {Kasper}, {Kumar}, {Phan}, \&
  {Velli}}]{Rauoafi:ApJ2023}
{Raouafi}, N.~E., {Stenborg}, G., {Seaton}, D.~B., {et~al.} 2023, \apj, 945, 28

\bibitem[{{Rappazzo} \& {Parker}(2013)}]{2013ApJ...773L...2R}
{Rappazzo}, A.~F. \& {Parker}, E.~N. 2013, Astrophys.\ J.\ Lett., 773, L2

\bibitem[{Rappazzo {et~al.}(2008)Rappazzo, Velli, Einaudi, \&
  Dahlburg}]{Rappazzo:2008vl}
Rappazzo, F., Velli, M., Einaudi, G., \& Dahlburg, R.~B. 2008, Astrophys.\ J.,
  677, 1348

\bibitem[{{Rosseland}(1924)}]{Rosseland_1924}
{Rosseland}, S. 1924, MNRAS, 84, 720

\bibitem[{Scudder(1992{\natexlab{a}})}]{Scudder1992a}
Scudder, J.~D. 1992{\natexlab{a}}, Astrophys.\ J., 398, 299

\bibitem[{Scudder(1992{\natexlab{b}})}]{Scudder1992b}
Scudder, J.~D. 1992{\natexlab{b}}, Astrophys.\ J., 398, 319

\bibitem[{{Teriaca} {et~al.}(2004){Teriaca}, {Banerjee}, {Falchi}, {Doyle}, \&
  {Madjarska}}]{Teriaca:2004wy}
{Teriaca}, L., {Banerjee}, D., {Falchi}, A., {Doyle}, J.~G., \& {Madjarska},
  M.~S. 2004, Astron.\ Astrophys., 427, 1065

\bibitem[{{Tiwari} {et~al.}(2019){Tiwari}, {Panesar}, {Moore}, {De Pontieu},
  {Winebarger}, {Golub}, {Savage}, {Rachmeler}, {Kobayashi}, {Testa}, {Warren},
  {Brooks}, {Cirtain}, {McKenzie}, {Morton}, {Peter}, \&
  {Walsh}}]{Tiwari:2019us}
{Tiwari}, S.~K., {Panesar}, N.~K., {Moore}, R.~L., {et~al.} 2019, Astrophys.\
  J., 887, 56

\bibitem[{{Wilmot-Smith}(2015)}]{2015RSPTA.37340265W}
{Wilmot-Smith}, A.~L. 2015, Philosophical Transactions of the Royal Society of
  London Series A, 373, 20140265

\bibitem[{{Yang, S. H.} {et~al.}(2009){Yang, S. H.}, {Zhang, J.}, {Jin, C. L.},
  {Li, L. P.}, \& {Duan, H. Y.}}]{observedtemperature}
{Yang, S. H.}, {Zhang, J.}, {Jin, C. L.}, {Li, L. P.}, \& {Duan, H. Y.} 2009,
  Astron.\ Astrophys., 501, 745

\bibitem[{{Zouganelis} {et~al.}(2004){Zouganelis}, {Maksimovic},
  {Meyer-Vernet}, {Lamy}, \& {Issautier}}]{Zouganelis2004}
{Zouganelis}, I., {Maksimovic}, M., {Meyer-Vernet}, N., {Lamy}, H., \&
  {Issautier}, K. 2004, Astrophys.\ J., 606, 542

\end{thebibliography}

\appendix

\section{Application to high-mass main-sequence stars ($M>1.5\, M_{\odot}$)}
\label{sec:main_sequence_stars_HMS}
In the previous Sections we have considered the case of low-mass main-sequence stars, i.e., stars whose mass $M$ is smaller than $1.5\, M_{\odot}$. As discussed in the Introduction, the case of larger-mass main-sequence stars is different because these stars lack a convective region below the photosphere and therefore should not have a solar-like magnetic field able to support a stationary corona: an X-ray emission from these stars, which is indeed detected, is not attributed to a Sun-like coronal activity. Moreover, the model we have applied to low-mass main-sequence stars assumes from the outset the presence of stationary coronal loops, and therefore may not be applicable as such to cases where there are no stationary magnetic configurations able to support these loops. 
This notwithstanding, it is interesting to study what are the predictions of the model in the case $M > 1.5\, M_{\odot}$, because it turns out that density and temperature profile exhibit a richer behaviour than in the low-mass case, and   
this may give some hints to possible generalizations of the model able to encompass also the case of massive stars.  

\begin{figure}[b]
    \centering
    \includegraphics[width=0.99\columnwidth]{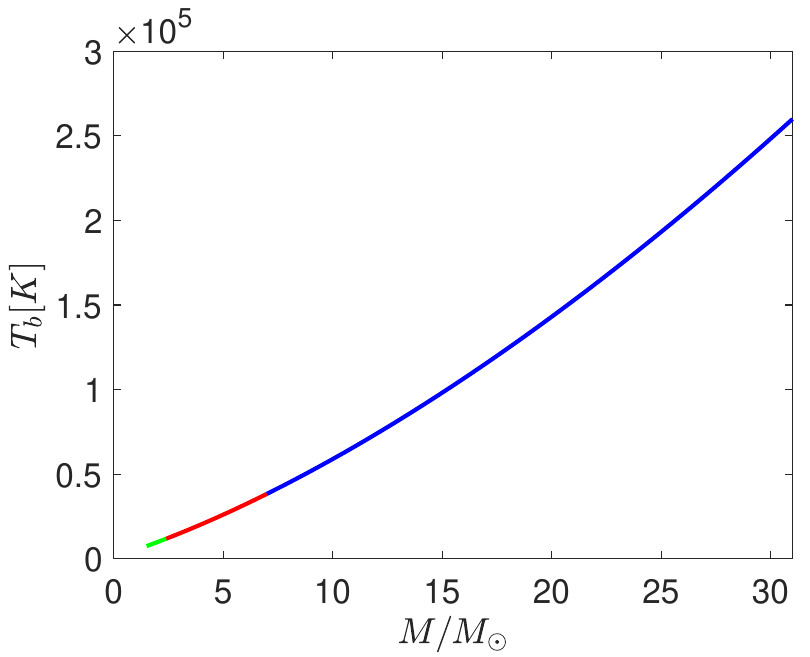}
    \caption{
    Plot of the thermostat (chromosphere) temperature, $T_b$ in Kelvin, as a function of the mass M (in units of solar mass $M_{\odot}$). The green curve corresponds to the mass interval $1.5 < M/M_{\odot} < 2.4$, the red curve to $2.4 < M/M_{\odot} < 7$, the blue curve to $7 < M/M_{\odot} < 31$.}
    \label{fig:TbaseHMS}
\end{figure}

\begin{figure}
    \centering
    \includegraphics[width=0.99\columnwidth]{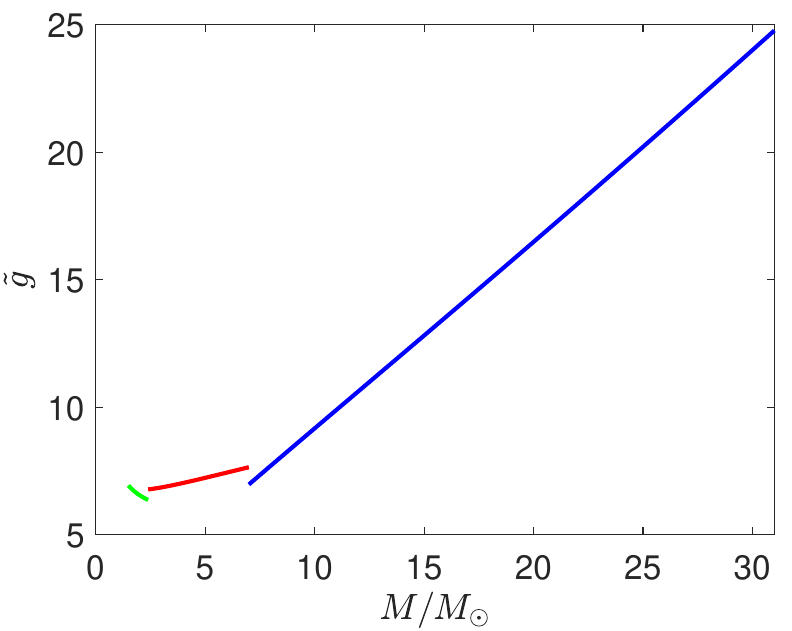}
    \caption{As in Fig.\ \ref{fig:TbaseHMS}, for the stratification parameter $\tilde{g}$ as a function of the mass $M$ (in units of solar mass $M_{\odot}$) for a fixed value of $z_t$ such that $z_t/R = 1/35$.}
    \label{fig:gtildeHMS}
\end{figure}

\begin{figure}[b]
    \centering
    \includegraphics[width=0.99\columnwidth]{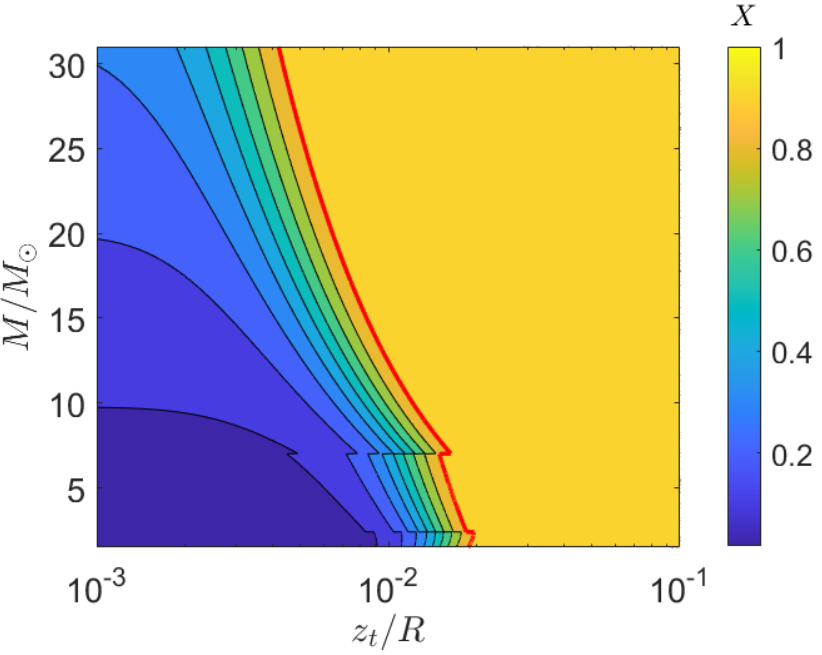}
    \caption{Contour plot of $X$ as defined in Eq.\ \eqref{parameterX}, computed for high-mass stars (i.e., $M \in [1.5 \, M_{\odot},31\, M_{\odot}]$). $X$ is plotted as a function of the star mass $M$ in units of solar mass $M_{\odot}$ and of the top height of the loop $z_t$ scaled by the star radius $R$. As in Fig.\ \ref{fig:ContoursmallM}, the red line corresponds to $X=0.9$, the threshold separating the regime without a Sun-like corona ($X < 0.9$, bluish colours) from that where there is a Sun-like corona ($X > 0.9$, yellowish colours).}
    \label{fig:ContourhighM}
\end{figure}

\begin{figure}
    \centering
    \includegraphics[width=0.99\columnwidth]{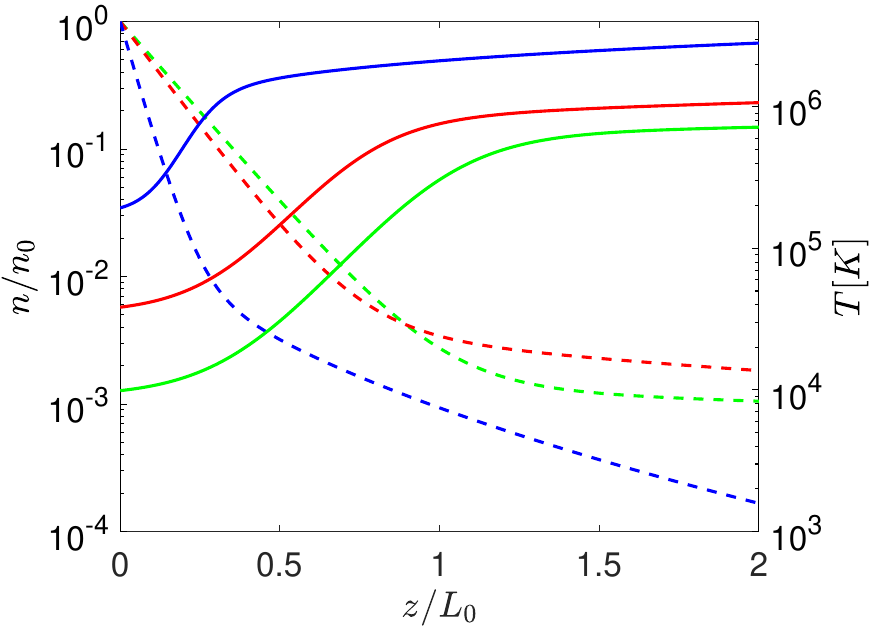}
    \caption{Density $n$ (in units of the density at the base of the loop, $n_0 = n(z = 0)$) and temperature $T$ (in Kelvin) as a function of the height $z$ within the loop, scaled by $L_0 = L/\pi$ (where $2L$ is the loop length), for some values of the mass in the range $1.5 < M/M_{\odot} < 31$. As in Fig.\ \ref{fig:TvsnMsmallMsun}, here we choose $z_t/R = 1/35$, such that $X > 0.9$ for all the values of $M$ and there always is a Sun-like corona. Solid lines correspond to temperatures and dashed lines to densities. The green curves are computed for $M = 2\, M_{\odot}$, the red ones for $M = 7 \, M_{\odot}$, the blue ones for $M = 25 \, M_{\odot}$ .
    }
    \label{fig:TvsnMlargeMsun}
\end{figure}

In close analogy to what we have done in Sec.\ \ref{sec:main_sequence_stars_LMS}, we assume that the relation between the surface temperature $T$ of a star and its mass $M$ is also applicable to the chromospheric temperature, which we identify with the reference temperature of the thermostat $T_b$. Using the scaling relations valid for high-mass main-sequence stars given by \cite{Eker2018} for $1.5 \, M_{\odot} \le M \le 31 \,  M_{\odot}$ we find that $T_b$ is an increasing function of $M$, as for low-mass stars, although in this case the range of values of $T_b$ is larger because the range of masses is larger, as shown in Fig.\ \ref{fig:TbaseHMS}, where $T_b$ is plotted against $M/M_{\odot}$.  
Using the above-mentioned relation we can write the dimensionless parameter $\Delta \tilde{T}$ for a generic star in terms of $M$ using its definition given in Eq.\ \eqref{eq:Delta_tilde_T} and fixing the mean value of the temperature increments to $\langle \Delta T \rangle = 9 \times 10^5$ K. 
Using also the relation between the radius $R$ and the mass $M$, again given by \cite{Eker2018}, we can express $\tilde{g}$ in terms of $M$ and of the loop height $z_{t}$. The result is shown in Fig.\ \ref{fig:gtildeHMS}, for the same fixed value of $z_t$ used in Sec.\ \ref{sec:main_sequence_stars_LMS}: at variance with the case of low-mass stars, $\tilde{g}$ is now an increasing function of $M$.  
This result implies that in the case of high-mass main-sequence stars a Sun-like corona is more readily achievable as the mass increases, as shown in Fig.\ \ref{fig:ContourhighM}, where the quantity $X$, as defined in Eq.\ \eqref{parameterX}, is plotted as a function of $z_{t}/R$ and $M/M_{\odot}$, analogously to what done in Fig.\ \ref{fig:ContoursmallM} for the case of low-mass stars. Apart from this, Fig.\ \ref{fig:ContourhighM} shows that the model would predict a Sun-like corona even in the high-mass case and for all the range of masses considered.   
More interesting differences with respect to the low-mass case show up when we look directly at the temperature and density profiles, reported in Fig.\ \ref{fig:TvsnMlargeMsun} for three choices of $M$ in the high-mass range and for $z_t = R/35$, a choice ensuring that there is a Sun-like corona. Comparing Fig.\ \ref{fig:TvsnMlargeMsun} with Fig.\ \ref{fig:TvsnMsmallMsun} we first notice that not only the temperature at the base of the loop increases with $M$ (as it happened for low-mas stars, and the larger variation of $T$ at $z = 0$ with respect to the low-mass case is only due to the larger variation of mass, as shown in Fig.\ \ref{fig:TbaseHMS}), but also the temperature at the top of the loop increases, at variance with the low-mass case. For high-mass stars, the temperature at the top of the loop increases from $7 \times 10^5$ K to $3 \times 10^6$ K when $M$ varies from $2\, M_{\odot}$ to $25 \, M_{\odot}$, while it remained around $10^6$ K in the low-mass case: this means that for high masses the temperature in the upper corona is no longer totally specified by the value of $\langle \Delta T \rangle$ (remember that the latter quantity has been fixed to $\langle \Delta T \rangle = 9 \times 10^5$ K). At the same time, if it is true that  $T_{\rm b}$ increases with $M$ up to values of about $2.5\times 10^5~{\rm K}$ for $M/M_{\odot}=30$, such contribution is not enough to explain such growth. 
The second difference between the high-mass and low-mass case is that in the latter the density profiles are all very similar to each other and depend monotonically on $M$, the density drop being smaller for heavier stars (see Fig.\ \ref{fig:TvsnMsmallMsun}), while the density profiles of high-mass stars reported in Fig.\ \ref{fig:TvsnMlargeMsun} are different from each other and their features do not monotonically depend on $M$: the density drop in the upper corona is smaller for the intermediate mass case, and the density profile in the upper corona is steeper in the larger mass case with respect to the other two masses considered. A third difference is that the width of the transition region gets smaller when increasing $M$; but if the latter feature is again a consequence of the already noted fact that $\tilde{g}$ increases with $M$ at variance with the low-mass case, the previous two differences are less easily explained.
\begin{figure}
    \centering
    \includegraphics[width=0.99\columnwidth]{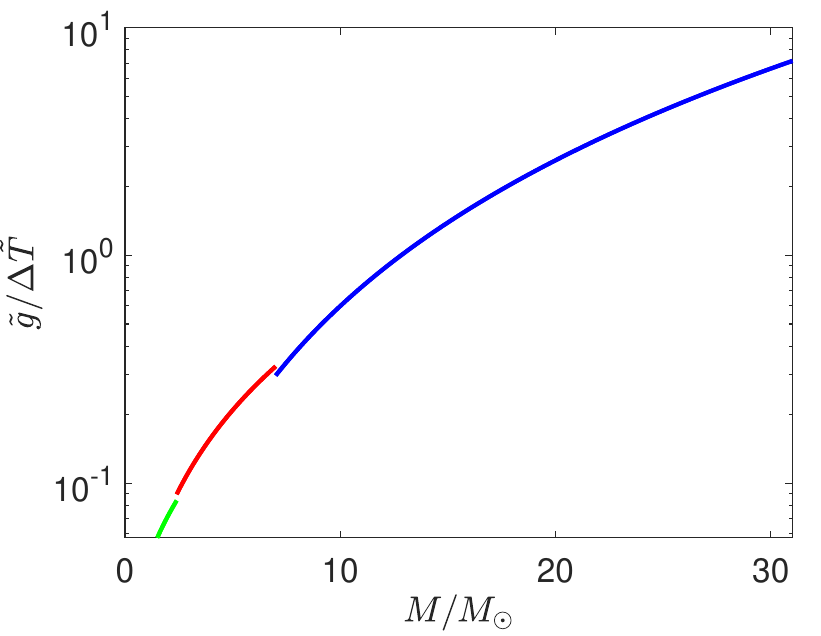}
    \caption{The stratification parameter $\tilde{g}$ scaled by $\Delta{\tilde{T}}$ as function of the mass M (in units of solar mass $M_{\odot}$) for a fixed value of $z_t$ that is $z_t/R = 1/35$. Colours are as in Fig.\ \ref{fig:TbaseHMS}.}
    \label{fig:gtildeDeltaT}
\end{figure}
\begin{figure}
    \centering
    \includegraphics[width=0.99\columnwidth]{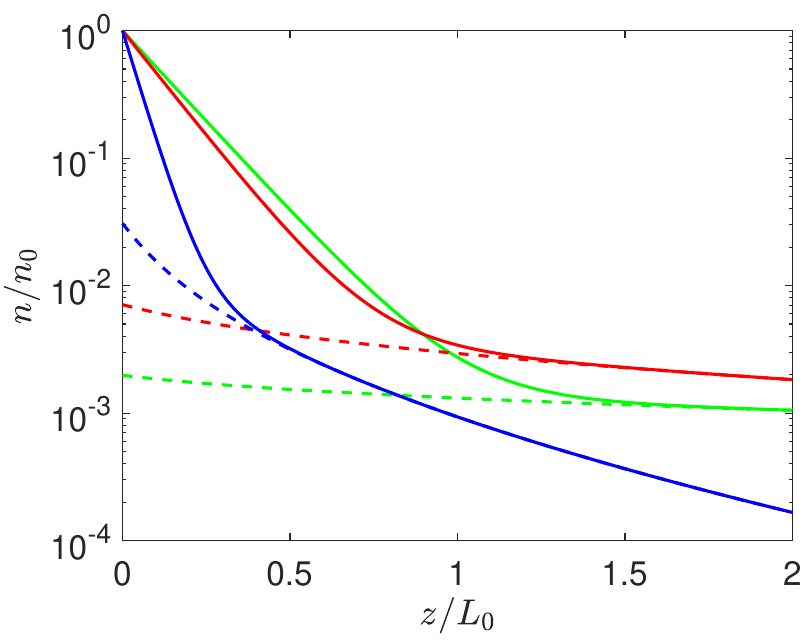}
    \caption{Comparison between the density profiles already shown in Fig.\ \ref{fig:TvsnMlargeMsun} (solid curves) and those obtained by considering only the ``hot'' population, that is, by retaining only the last term in the Eq.\ \eqref{eq:falphastationary}. Colours are as in Fig.\ \ref{fig:TbaseHMS}.}
    \label{fig:DensityvsDensitymulti}
\end{figure}
In order to explain these differences we report in Fig.~\ref{fig:gtildeDeltaT} the following parameter
\begin{equation}\label{eq:stratcoronalpart}
    \frac{\tilde{g}}{\Delta \tilde{T}}= \frac{g(m_e+m_p)L}{2\pi k_B \langle \Delta T \rangle}\,, 
\end{equation}
which is a measure of the strength of the gravitational bond (gravitational energy) with respect to the typical thermal energy of a coronal particle. As $M$ increases such quantity grows and it is harder for particles to reach the top of the coronal loop. As a consequence, the velocity filtration mechanism is stronger and it produces a higher temperature in corona, much larger than the contribution of the cold population. 
The behaviour of the density can be still understood in terms of the parameter in Eq.~\ref{eq:stratcoronalpart}. In Fig.~\ref{fig:DensityvsDensitymulti} we plot the density profiles already shown in Fig.~\ref{fig:TvsnMlargeMsun} (solid curves), together with the contribution to the density of the ``hot'' population (dashed curves). For relatively small masses the contribution to the total density of such population is small and dominates only in the upper part of the loop, decreasing smoothly because of the small value of $\tilde{g}/\Delta \tilde{T}$. For intermediate values of the masses the contribution of the ``hot'' population to the total density increases and we observe an increase in the density in the top of the loop. When the mass of the star becomes very high the contribution of the ``hot'' population to the total density still increases but now the gravitational bond is so strong that the density in the loop decreases very rapidly and the density at the top of the loop becomes smaller than those observed for small and intermediate values of the star mass.    

\end{document}